\documentclass[%
reprint,
amsmath,amssymb,
aps,
]{revtex4-2}
\usepackage{float}
\usepackage{graphicx}
\usepackage{dcolumn}
\usepackage{bm}

\begin{document}
	\preprint{APS/123-QED}
\title{Quantum Coherence in Reflected and Refracted Beams: A Van Cittert-Zernike Approach}
\author{Yuetao Chen$^{1}$}
\author{Gaiqing Chen $^{1}$}
\author{Jin Wang$^{1}$}
\author{Qiang Ma$^{1}$}
\author{Shoukang Chang$^{1}$}
\author{Shaoyan Gao$^{1}$}
\thanks{Corresponding author. gaosy@xjtu.edu.cn}
\affiliation{$^{{\small 1}}$\textit{MOE Key Laboratory for Nonequilibrium Synthesis and
		Modulation of Condensed Matter, Shaanxi Province Key Laboratory of Quantum
		Information and Quantum Optoelectronic Devices, School of Physics, Xi'an
		Jiaotong University, 710049, China}}

\begin{abstract}
Recent advances in quantum optics have highlighted the critical role of spatial propagation in controlling the quantum coherence of light beams. However, the evolution of quantum coherence for light beams undergoing fundamental optical processes at dielectric interfaces remains unexplored. Furthermore, manipulating multiphoton correlations typically requires complex interactions that challenge few-photon level implementation. Here, we introduce a quantum van Cittert–Zernike theorem for light beams, describing how their coherence–polarization properties are influenced by reflection and refraction, as well as how these properties evolve upon subsequent propagation. Our work demonstrates that the quantum statistics of photonic systems can be controllably modified through the inherent polarization coupling arising from reflection and refraction at an interface, without relying on conventional light-matter interactions. Our approach reveals regimes where thermal light can exhibit sub-Poissonian statistics with fluctuations below the shot-noise level through post-selected measurements, and this statistical property can be tuned by the incident angle. Remarkably, this quantum statistical modification is governed by a scaling law linking beam collimation to far-field thermalization. Our work establishes a robust, decoherence-avoiding mechanism for quantum state control, advancing the fundamental understanding of coherence in quantum optics and opening new avenues for applications in quantum information and metrology.

\end{abstract}

\maketitle

\section{Introduction}
The van Cittert–Zernike theorem stands as a cornerstone of optical physics\cite{vanCittert1934,Zernike1938}, providing the theoretical framework to describe how the coherence properties of light evolve during propagation\cite{vanCittert1934,Zernike1938,Born2013,Wolf1954}. This foundational principle has spurred decades of research into the dynamics of spatial, temporal, spectral, and polarization coherence across a wide range of optical beams\cite{Dorrer2004,Gori2000,Cai2020,Cai2012}. In classical optics, its application has been pivotal for advancing techniques in optical sensing, metrology, and astronomical interferometry\cite{Carozzi2009,Batarseh2018,Barakat2000}. More recently, the theorem’s implications have extended into the quantum realm, prompting investigations into its role for quantum systems\cite{Barbosa1996,Saleh2005,Fabre2017,PhysRevLett.123.143604,Barrachina2020,Khabiboulline2019}. Significant work has focused on understanding how spatial coherence evolves in biphoton states\cite{Barbosa1996,Saleh2005,PhysRevLett.123.143604,Khabiboulline2019,Reichert2017,Defienne2019}, with extensions of the theorem developed to quantify spatial entanglement in photon pairs generated via parametric down-conversion\cite{Barbosa1996,Saleh2005,Qian2018,Eberly2016}. Insights from these studies on the propagation of spatial coherence and entanglement have proven essential for advancing quantum technologies, including metrology, spectroscopy, imaging, and lithography\cite{Barbosa1996,Saleh2005,PhysRevLett.123.143604,Khabiboulline2019,Bhusal2022,LeonMontiel2013,You2021,OBrien2009,Wen2013}. Nevertheless, previous investigations have overlooked the role of the quantum van Cittert-Zernike theorem in governing the evolution of multiphoton correlations and photon statistics for beams undergoing reflection and refraction at dielectric interfaces\cite{Barbosa1996,Saleh2005,Fabre2017,PhysRevLett.123.143604,Barrachina2020,Khabiboulline2019,Reichert2017,Defienne2019,Qian2018,Eberly2016,Bhusal2022,LeonMontiel2013,You2021,OBrien2009,Wen2013}.

The reflection and refraction of a plane wave at an interface are well described by the Fresnel equations and Snell’s law \cite{Born1999}. For a light beam, however, these laws require modification because the beam possesses a distributed angular spectrum \cite{DErrico17}. This distributed angular spectrum is responsible for the in-plane and out-of-plane spatial displacements—the Goos–Hänchen (GH) and Imbert–Fedorov (IF) shifts \cite{chen2023beam,chen2025estimation},  which have been confirmed in experiments \cite{goswami2016optimized,zha2024optical,joshua2025observation}. The key point is that an adequate description of beam reflection or refraction essentially involves performing rotational transformations for each plane-wave component in the angular spectrum \cite{bliokh}. Interestingly, this rotational transformations give rise to off-diagonal elements in the Jones matrices for reflection/refraction, directly coupling the horizontal and vertical polarization states as a natural feature of beam propagation at an interface. Such coupling, which avoids the need for complex light-matter interactions that are challenging to implement at the few-photon level \cite{dell2006multiphoton,venkataraman2013phase}, provides a practical means to manipulate beam coherence-polarization in both single- \cite{gori2000use} and multi-photon systems \cite{you2023multiphoton} via polarization gratings, thereby circumventing the unavoidable decoherence typical of nonclassical systems interacting with realistic environments \cite{yu2009sudden}. Beyond its potential for manipulating beam coherence-polarization, the polarization coupling arising from the reflection/refraction process also influences a key component in quantum optics: the beam splitter (BS) \cite{scully1997quantum,agarwal2012quantum}. Since a BS fundamentally operates via the partial reflection and transmission of a light beam , this inherently weak yet non-negligible coupling effect merits consideration in quantum optical setups.

Therefore, to explore the potential of polarization coupling in beam reflection/refraction for manipulating coherence-polarization and to investigate its influence on BS, we study the quantum coherence of reflected and transmitted beams based on the quantum van Cittert–Zernike theorem. In our model, the dielectric interface is illuminated by two incident Gaussian beams with equal angles of incidence, simultaneously undergoing reflection and transmission, which effectively simulates the action of a BS. In contrast to established paradigms in quantum optics \cite{Barbosa1996,Saleh2005,Fabre2017,PhysRevLett.123.143604,Barrachina2020,Khabiboulline2019,Reichert2017,Defienne2019,Qian2018,Eberly2016,Bhusal2022,LeonMontiel2013,You2021,OBrien2009,Wen2013}, we show that the quantum statistics of reflected and transmitted beams can be controllably altered during free-space propagation, without relying on conventional light-matter interactions. These correspond to optical processes that occur without photon absorption or emission \cite{Allen1987}. Interestingly, our work suggests regimes in which thermal light could, in principle, give rise to fluctuations below the shot-noise level. This primarily stems from the polarization coupling inherent in beam reflection and refraction, a natural phenomenon governed by rotational transformations \cite{bliokh}. Notably, this coupling can be precisely controlled via the angle of incidence, thereby establishing it as a novel, naturally accessible degree of freedom for manipulating quantum coherence. Moreover, our work implies that for a BS, the quantum coherence of a poorly collimated Gaussian beam is more susceptible to this polarization-coupling effect. In other words, for a well-collimated Gaussian beam, the influence of such coupling on quantum coherence is essentially negligible. Our work establishes an alternative approach—specifically via the degree of freedom offered by the angle of incidence—to control the quantum coherence of a light beam through its reflection and refraction, which holds potential for applications in areas such as quantum information and quantum metrology.

This paper is organized as follows: In Sec. \ref{Model}, we introduce a model for beam reflection and refraction and employ the quantum van Cittert–Zernike theorem to describe the propagation of coherence–polarization in the reflected and transmitted beams. In Sec. \ref{RESULTS AND DISCUSSION}, we first examine how the position of the detector on the screen and the screen distance influence the beam’s coherence–polarization. We then study the effect of the incidence angle on coherence–polarization, and finally discuss the role of Gaussian-beam collimation in the thermalization of the optical field.  Finally, we present the conclusions of our results in Sec. \ref{CONCLUSION}.
\begin{figure*}[ht]
	\label{Fig.1} \centering \includegraphics[width=1.3\columnwidth]{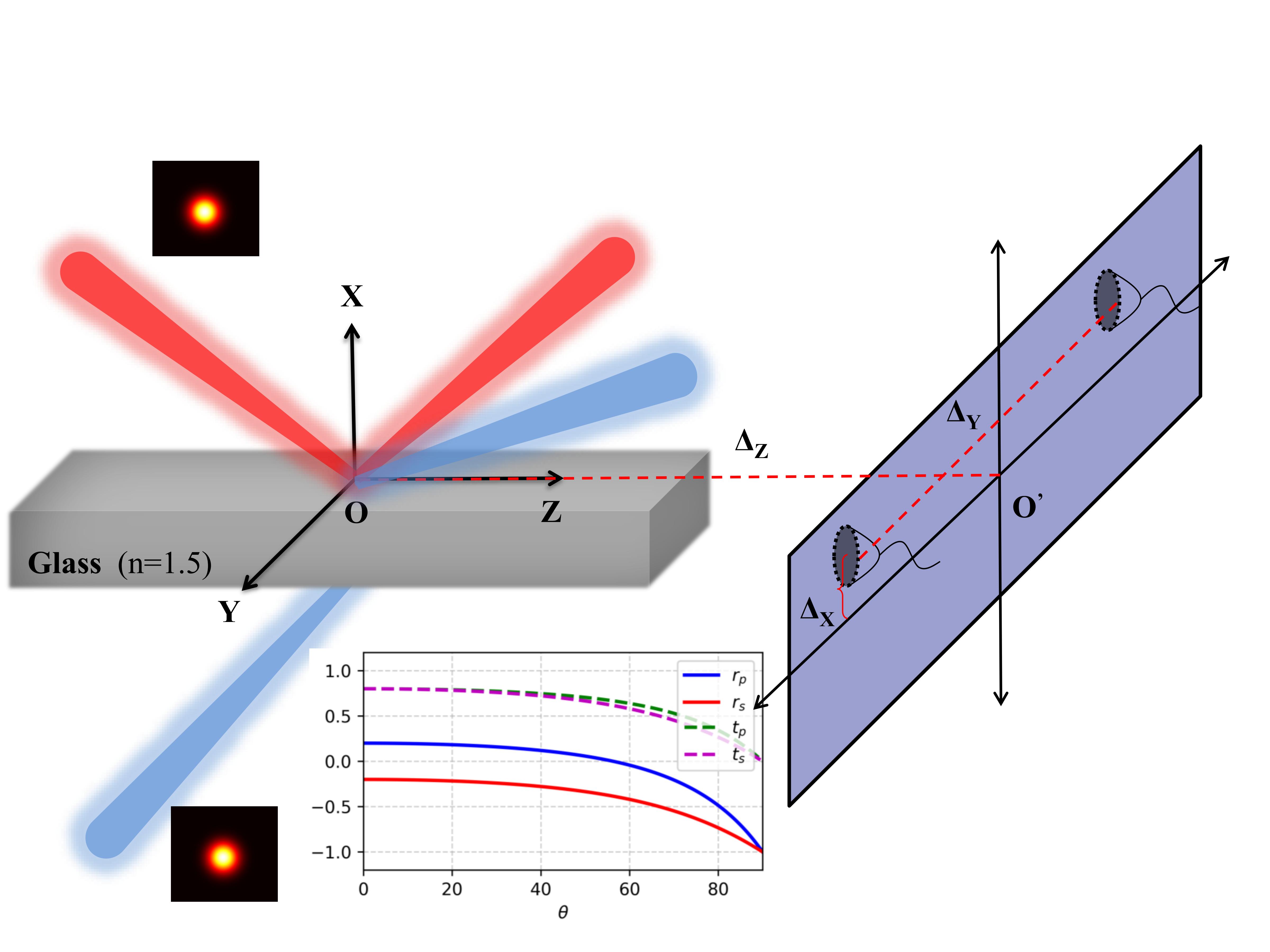}
	\caption{The proposed setup for investigating quantum van Cittert-Zernike theorem for reflection and refraction of light beams. In this setup, both the incident blue and yellow beams are Gaussian beams. The inset shows the intensity distribution of their transverse cross-section. The incident beams interact with a dielectric medium of refractive index $n=1.5$, whose reflection coefficients ($r_p$, $r_s$) and transmission coefficients ($t_p$, $t_s$) are illustrated in an accompanying plot. After reflection and refraction, the Gaussian beams propagate to a screen located at a distance of $\Delta_Z$. Two detectors are symmetrically placed on the screen with a horizontal separation of $\Delta_Y$ and are positioned at a vertical height of $\Delta_X$ from the base of the screen.}
\end{figure*}
\section{Model}\label{Model}
We investigate the propagation of multiphoton two-mode correlations in the setup shown in Fig. 1, demonstrating the quantum van Cittert–Zernike theorem. We consider two thermal, spatially incoherent, unpolarized Lagrange-Gauss beams that interacts with a glass of refractive index $n=1.5$. The two beams are reflected (yellow) and transmitted (blue) by the glass, respectively, then propagate in the far field to a screen at distance $\Delta_Z$, where they are measured by two detectors separated by a distance $\Delta_Y$ with a vertical height difference of $\Delta_X$. By post-selecting the detected intensities, we determine the correlations between different beam modes. The correlations here is two-photon four-point correlation matrix, which can be written as
\begin{widetext} 
\begin{equation}\label{Eq_1}
g^{(2)}_{\alpha\beta\alpha^{\prime}\beta^{\prime}}\left(\boldsymbol{r}_1,\boldsymbol{r}_2;\boldsymbol{r}_3,\boldsymbol{r}_4,z\right)=\left\langle j_{\alpha\beta}\left(\boldsymbol{r}_1,\boldsymbol{r}_2,z\right)\otimes j_{\alpha^{\prime}\beta^{\prime}}\left(\boldsymbol{r}_3,\boldsymbol{r}_4,z\right)\right\rangle,
\end{equation}
\end{widetext} 
where $\langle j_{\alpha(\alpha^{\prime})\beta(\beta^{\prime})}\left(\boldsymbol{r}_{1(3)},\boldsymbol{r}_{2(4)},z\right)\rangle=\left\langle E_{\alpha(\alpha^{\prime})}^{(-)}\left(\boldsymbol{r}_{1(3)},z;t\right)E_{\beta(\beta^{\prime})}^{(+)}\left(\boldsymbol{r}_{2(4)},z;t\right)\right\rangle$ with $\alpha,\beta,\alpha^{\prime},\beta^{\prime}=H,V$. The time average $\langle\cdots\rangle$ contains the negative- and positive-frequency field components $E_{\alpha}^{(\mp)}(\mathbf{r},z;t)$ for the $\alpha$ polarization. $z$ and $r$ represent the propagation distance and the transverse coordinate in the beam cross-section, respectively. Following previous authors \cite{four-pointcorrelation1,four-pointcorrelation2,four-pointcorrelation3}, a unpolarized, spatially incoherent source is characterized by a four-point correlation matrix can be written as 
\begin{widetext}
	\begin{equation}\label{Eq_2}
		\begin{aligned}
			&g^{(2),in}\left(x_1,y_1,x_2,y_2;x_3,y_3,x_4,y_4,0\right)  = \lambda^4 I\left(x_1,y_1\right) I\left(x_3,y_3\right)  F\times \mathbf{I}^1_{2 \times 2} \otimes \mathbf{I}^2_{2 \times 2}, \\
			&F = \left[\delta\left(y_2-y_1\right)\delta\left(x_2-x_1\right)\delta\left(y_3-y_4\right)\delta\left(x_3-x_4\right) + \delta\left(y_2-y_3\right)\delta\left(x_2-x_3\right)\delta\left(y_1-y_4\right)\delta\left(x_1-x_4\right)\right]
		\end{aligned}
	\end{equation}
\end{widetext}
where $I(r)$ is the beam intensity that depends on position $r$, and $\lambda$ is the wavelength of the beam. The two unit matrices $\mathbf{I}^{1,2}_{2 \times 2}$ describe the polarization density matrices of the two unpolarized light beams. It's worth noting that the second sum of delta functions in Eq. (\ref{Eq_2}) stems from the indistinguishability of photon. 

We consider that both sources are in the Gaussian mode, denoted by the yellow and blue beams in Fig. 1. Thus, we have ${I}(x,y)=2/(\pi{w}_{0}^{2})\exp(-(x^2+y^2)/{w}_{0}^{2})$, and its angular spectrum takes the form $I(k_x^i,k_y^i)=\frac{2w_0^2}{\pi}\exp\left[-\frac{w_0^2}{2}({k_x^i}^2+{k_y^i}^2)\right]$, with $k_x^i$ and $k_y^i$ being the transverse momenta of incident beam in the $x$ and $y$ directions. Notably, the two incident Gaussian beams interact with the glass, with one beam reflected and the other transmitted, thus effectively simulating a beam splitter (BS)—a ubiquitous component in quantum optics experiments. 

The reflection and transmission of a beam cannot be fully described by Snell's law alone, since an accurate treatment first requires a Fourier expansion of the incident beam into an angular spectrum that incorporates wavevectors outside the $(x,z)$ plane of incidence. Then, each angular spectrum component with a distinct wavevector must be individually rotational transformed among the three coordinate frames \cite{bliokh}. It yields the complete transformation and the Jones matrix relating the incident and secondary fields in the beam coordinate frames, thereby offering a full description of the beam's reflection and transmission by the medium:$|\mathbf{\tilde{E}}^{a=r,t})_{\perp}=\mathbf{\hat{J}}_{\perp}^{a=r,t}|\mathbf{\tilde{E}})_{\perp}$ where
\vspace{0.6em}
\begin{small}
\begin{equation}\label{Eq_3}\begin{aligned}\hat{J}_{\perp}^{r}\simeq\begin{pmatrix}r_p+\frac{\partial r_p}{\partial\theta}\frac{k_{x_1}^r}{k_1}&(r_p+r_s)\cot\theta\frac{k_{y_1}^r}{k_1}\\-(r_p+r_s)\cot\theta\frac{k_{y_1}^r}{k_1}&r_s+\frac{\partial r_s}{\partial\theta}\frac{k_{x_1}^r}{k_1}\end{pmatrix},\\\hat{J}_{\perp}^{t}\simeq\begin{pmatrix}t_p+\eta\frac{\partial t_p}{\partial\theta}\frac{k_{x_3}^t}{k_3}&(t_p-\eta t_s)\cot\theta\frac{k_{y_3}^t}{k_3}\\\left(\eta t_{p}-t_{s}\right)\cot\theta_{i}\frac{k_{y_3}^t}{k_3}&t_s+\eta\frac{\partial t_s}{\partial\theta}\frac{k_{x_3}^t}{k_3}\end{pmatrix},\end{aligned}\end{equation}
\end{small}
\vspace{0.6em}
where $\theta$ is the incident angle, $\eta=\cos\theta_{t}/\cos\theta$, with $\theta_{t}$ is the angle of refraction. $k_{x_1,y_1}^r$ and $k_{x_3,y_3}^t$ denote the transverse momenta along the $x$ and $y$ directions for the reflected and transmitted beams, respectively. Then, analogous to the action of a polarization grating on an unpolarized beam \cite{polarizer}, the process of reflection and transmission of an unpolarized beam by the medium is represented by $\mathbf{\hat{J}}_{\perp}^{r\dagger}\mathbf{I}^1_{2 \times 2}\mathbf{\hat{J}}_{\perp}^{r}$ and $\mathbf{\hat{J}}_{\perp}^{t\dagger}\mathbf{I}^2_{2 \times 2}\mathbf{\hat{J}}_{\perp}^{t}$, respectively. Consequently, the four-point correlation matrix of the two incident beams, after interaction with the medium, can be expressed as
\begin{small}
\begin{equation}\label{Eq_4}
\begin{aligned}
&g^{(2),out}\left(x_1,y_1,x_2,y_2;x_3,y_3,x_4,y_4,0\right)=\\&\int\int\int\int I(k_{x_1}^r,k_{y_1}^r)I(k_{x_3}^t,k_{y_3}^t)F\times\mathbf{\hat{J}}_{\perp}^{r\dagger}\mathbf{I}^1_{2 \times 2}\mathbf{\hat{J}}_{\perp}^{r}\otimes\mathbf{\hat{J}}_{\perp}^{t\dagger}\mathbf{I}^2_{2 \times 2}\mathbf{\hat{J}}_{\perp}^{t}\times\\ &e^{-i(k_{x_1}^{r}x_1+k_{x_3}^{t}x_3+k_{y_3}^{r}y_3+k_{y_3}^{t}y_3)}dk_{x_1}^{r}dk_{x_1}^{t}dk_{y_3}^{r}dk_{y_3}^{t}.
\end{aligned}
\end{equation}
\end{small}
After propagating a distance $\Delta_Z$ to the screen, the four-point correlation matrix is detected, and its expression is given by 	
\begin{widetext}
\begin{equation}\label{Eq_5}\begin{aligned}&g^{(2),out}\left(x'_1,y'_1,x'_2,y'_2;x'_3,y'_3,x'_4,y'_4,\Delta_Z\right)=\int\int\int\int\int\int\int\int g^{(2),out}\left(x_1,y_1,x_2,y_2;x_3,y_3,x_4,y_4,0\right)\times\\&K^{*}\left(x_{1},x'_{1},y_1,y'_1,\Delta_Z\right) K\left(x_{2},x'_{2},y_2,y'_2,\Delta_Z\right) K^{*}\left(x_{3},x'_{3},y_3,y'_3,\Delta_Z\right)K\left(x_{4},x'_{4},y_4,y'_4,\Delta_Z\right)dx_{1}dy_{1}dx_{2}dy_{2}dx_{3}dy_{3}dx_{4}dy_{4},\end{aligned}\end{equation}\end{widetext}
with the Fresnel propagation kernel defined by \cite{goodman2005introduction}
\begin{widetext}
\begin{equation}\label{Eq_6}
		K(x,x',y,y',z) = \frac{-i\exp(ik\Delta_Z)}{\lambda \Delta_Z} 
		\exp\left[ \frac{ik_x}{2\Delta_Z}(x-x')^2 + \frac{ik_y}{2\Delta_Z}(y-y')^2 \right].
\end{equation}
\end{widetext}
After imposing the detector-coordinate conditions $x'_2=x'_3,y'_2=y'_3$ and $x'_1=x'_4,y'_1=y'_4$ for the two-photon correlation at plane $z$,introducing the variables $x'_1-x'_2=\Delta_X,y'_1-y'_2=\Delta_Y$, and normalizing Eq. (\ref{Eq_5}) following \cite{gori1998}, we find 
 \begin{widetext}
 \begin{equation}\label{Eq_7}
 	\begin{aligned}
 		g^{(2),out}_{HVHV}&=-\cot^2{\theta}\frac{\Delta_{Y}^{r}}{\Delta_{Z}^{r}}\frac{r_{p}+r_{s}}{r_pr_s}[\frac{\Delta_{X}^{r}}{\Delta_{Z}^{r}}(r_{p}^{\prime}+r_{s}^{\prime})+r_{p}+r_{s}]\times\frac{\Delta_{Y}^{t}}{\Delta_{Z}^{t}}\frac{t_{p}-\eta t_{s}}{t_pt_s}[-\frac{\Delta_{X}^{t}}{\Delta_{Z}^{t}}(t_{p}^{\prime}+t_{s}^{\prime})+t_{p}+t_{s}]\\ &\times\exp\{-\frac{w_{0}^{2}k_{0}^{2}}{2}\left[\left(\frac{\Delta_{X}^{r}}{\Delta_{Z}^{r}}\right)^{2}+\left(\frac{\Delta_{Y}^{r}}{\Delta_{Z}^{r}}\right)^{2}+\left(\frac{\Delta_{X}^{t}}{\Delta_{Z}^{t}}\right)^{2}+\left(\frac{\Delta_{Y}^{t}}{\Delta_{Z}^{t}}\right)^{2}\right]\}\\&+\frac{\lambda^{4}\cot^2\theta(r_p + r_s)(t_p - \eta t_s)[(r'_p + r'_s)\Delta_S-ik_{0}(r_p + r_s)w_{0}^{2}\Delta_{IF}][(t'_p + t'_s)\eta \Delta_S+ik_{0}(t_p + t_s)w_{0}^{2}\Delta_{IF}]}{4\pi^2 w_{0}^{4}r_pr_st_pt_s},
 	\end{aligned}
 		\end{equation}
 \end{widetext}
 
 \begin{widetext}
	\begin{equation}\label{Eq_8}
		\begin{aligned}
			g^{(2),out}_{VVHH}&=\frac{(\frac{\Delta_{X}^{r}}{\Delta_{Z}^{r}}r_{p}^{\prime}+\pi r_p)^2-(\cot{\theta}\frac{\Delta_{Y}^{r}}{\Delta_{Z}^{r}}\pi (r_p+r_s))^2}{2\pi^2r_sr_s}\times\frac{(\cot{\theta}\pi\frac{\Delta_{Y}^{t}}{\Delta_{Z}^{t}})^2(t_{s}-\eta t_{p})(\eta t_{s}- t_{p})-(\pi t_s-\pi \frac{\Delta_{X}^{t}}{\Delta_{Z}^{t}}\eta t'_s)^2}{2\pi^2t_pt_p}\\ &\times\exp\{-\frac{w_{0}^{2}k_{0}^{2}}{2}\left[\left(\frac{\Delta_{X}^{r}}{\Delta_{Z}^{r}}\right)^{2}+\left(\frac{\Delta_{Y}^{r}}{\Delta_{Z}^{r}}\right)^{2}+\left(\frac{\Delta_{X}^{t}}{\Delta_{Z}^{t}}\right)^{2}+\left(\frac{\Delta_{Y}^{t}}{\Delta_{Z}^{t}}\right)^{2}\right]\}+1,
		\end{aligned}
	\end{equation}
\end{widetext}

 \begin{widetext}
	\begin{equation}\label{Eq_9}
		\begin{aligned}
			g^{(2),out}_{VVVV}&=\frac{(\frac{\Delta_{X}^{r}}{\Delta_{Z}^{r}}r_{p}^{\prime}+\pi r_p)^2-(\cot{\theta}\frac{\Delta_{Y}^{r}}{\Delta_{Z}^{r}}\pi (r_p+r_s))^2}{2\pi^2r_sr_s}\times\frac{(\cot{\theta}\pi\frac{\Delta_{Y}^{t}}{\Delta_{Z}^{t}})^2(t_{s}-\eta t_{p})(\eta t_{s}- t_{p})-(\pi t_p-\pi \frac{\Delta_{X}^{t}}{\Delta_{Z}^{t}}\eta t'_p)^2}{2\pi^2t_st_s}\\ &\times\exp\{-\frac{w_{0}^{2}k_{0}^{2}}{2}\left[\left(\frac{\Delta_{X}^{r}}{\Delta_{Z}^{r}}\right)^{2}+\left(\frac{\Delta_{Y}^{r}}{\Delta_{Z}^{r}}\right)^{2}+\left(\frac{\Delta_{X}^{t}}{\Delta_{Z}^{t}}\right)^{2}+\left(\frac{\Delta_{Y}^{t}}{\Delta_{Z}^{t}}\right)^{2}\right]\}+1,
		\end{aligned}
	\end{equation}
\end{widetext}

 \begin{widetext}
	\begin{equation}\label{Eq_10}
		\begin{aligned}
			g^{(2),out}_{HHVH}&=\frac{(\frac{\Delta_{X}^{r}}{\Delta_{Z}^{r}}r_{p}^{\prime}+\pi r_p)^2-(\cot{\theta}\frac{\Delta_{Y}^{r}}{\Delta_{Z}^{r}}\pi (r_p+r_s))^2}{2\pi^2r_pr_p}\times\frac{\Delta_{Y}^{t}}{\Delta_{Z}^{t}}\frac{t_{s}-\eta t_{p}}{t_pt_s}[\frac{\Delta_{X}^{t}}{\Delta_{Z}^{t}}(t_{p}^{\prime}+t_{s}^{\prime})-t_{p}-t_{s}]\\ &\times\exp\{-\frac{w_{0}^{2}k_{0}^{2}}{2}\left[\left(\frac{\Delta_{X}^{r}}{\Delta_{Z}^{r}}\right)^{2}+\left(\frac{\Delta_{Y}^{r}}{\Delta_{Z}^{r}}\right)^{2}+\left(\frac{\Delta_{X}^{t}}{\Delta_{Z}^{t}}\right)^{2}+\left(\frac{\Delta_{Y}^{t}}{\Delta_{Z}^{t}}\right)^{2}\right]\}\\&+\frac{\lambda^{4}\cot^2\theta(r_p + r_s)(t_s - \eta t_p)[(r'_p + r'_s)\Delta_S-ik_{0}(r_p + r_s)w_{0}^{2}\Delta_{IF}][(t'_p + t'_s)\eta \Delta_S+ik_{0}(t_p + t_s)w_{0}^{2}\Delta_{IF}]}{4\pi^2 w_{0}^{4}r_pr_st_pt_s},
		\end{aligned}
	\end{equation}
\end{widetext}
\begin{widetext}
	\begin{equation}\label{Eq_11}
		\begin{aligned}
			\Delta_Y=\frac{\iint_{-\infty}^{+\infty}{y}I(x,y)dxdy}{\iint_{-\infty}^{+\infty} I(x,y)dxdy},\Delta_S=\frac{\iint_{-\infty}^{+\infty}{xy}I(x,y)dxdy}{\iint_{-\infty}^{+\infty} I(x,y)dxdy}
		\end{aligned}
	\end{equation}
\end{widetext}
where $\Delta_{X}^{r} = [(\Delta_{X}\tan\theta + \Delta_{Z})\sin\theta - \Delta_{X}/\cos\theta]/(\Delta_{X}\sin\theta + \Delta_{Z}\cos\theta) $, $\Delta_{Y}^{r} = \Delta_{Y}/(\Delta_{X}\sin\theta + \Delta_{Z}\cos\theta) $, $\Delta_{Z}^{r} = \Delta_{X}\sin\theta + \Delta_{Z}\cos\theta $, $\Delta_{X}^{r} = [(\Delta_{X}\tan\theta_t + \Delta_{Z})\sin\theta_t - \Delta_{X}/\cos\theta_t]/(\Delta_{X}\sin\theta_t + \Delta_{Z}\cos\theta_t) $, $\Delta_{Y}^{r} = \Delta_{Y}/(\Delta_{X}\sin\theta_t + \Delta_{Z}\cos\theta_t) $, $\Delta_{Z}^{r} = \Delta_{X}\sin\theta_t + \Delta_{Z}\cos\theta_t $, based on geometrical considerations. $'$ is defined as the derivative with respect to the incidence angle $\theta$. Note that the additive term +1 in Eqs. (\ref{Eq_8}) and (\ref{Eq_9}) stems from the indistinguishability term in Eq. (\ref{Eq_2}). Eqs. (\ref{Eq_7}) and (\ref{Eq_10}) show that the multi-photon indistinguishability contribution to the off-diagonal correlations depends on the IF shift $\Delta_{IF}$ and shift area $\Delta_S$, which quantify beam aberration \cite{dennis2012topological,bliokh,chen2025estimation}. Thus, for a fundamental Gaussian beam without aberration, this contribution vanishes, as is directly confirmed by the integrals in Eq. (\ref{Eq_11}). Consequently, the second-order coherence $|g^{(2),out}|$ depends only on the ratio of the beam waist $w_0$ to the wavelength $\lambda$, as seen in the exponential factors of Eqs. (7)-(10). According to Eqs. (\ref{Eq_7}-\ref{Eq_10}), $g^{(2),out}$ depends on the distance between the detectors, $\Delta_Y$, the screen position $\Delta_Z$, and the vertical detector coordinate $\Delta_X$. Moreover, in contrast to schemes based on polarization gratings \cite{you2023multiphoton}, $g^{(2),out}$ here also depends on the angle of incidence, $\theta$. Given that Eqs. (\ref{Eq_7}-\ref{Eq_10}) describes any incoherent unpolarized state, we focus on two-mode thermal states to leverage their well-established statistical properties \cite{statisticalproperties1,statisticalproperties2}.
\section{RESULTS AND DISCUSSION}\label{RESULTS AND DISCUSSION}
Fig. 2 displays the normalized second-order coherence $|g^{(2),out}|$ as  functions of the ratio of separation $\Delta_Y$ to distance $\Delta_Z$ for multiple polarization-based measurement configurations. Each curve corresponds to a distinct post-selected measurement, revealing how spatial correlations evolve with increasing detector displacement.
\begin{figure}[ht]
	\label{Fig.2} \centering \includegraphics[width=1.0\columnwidth]{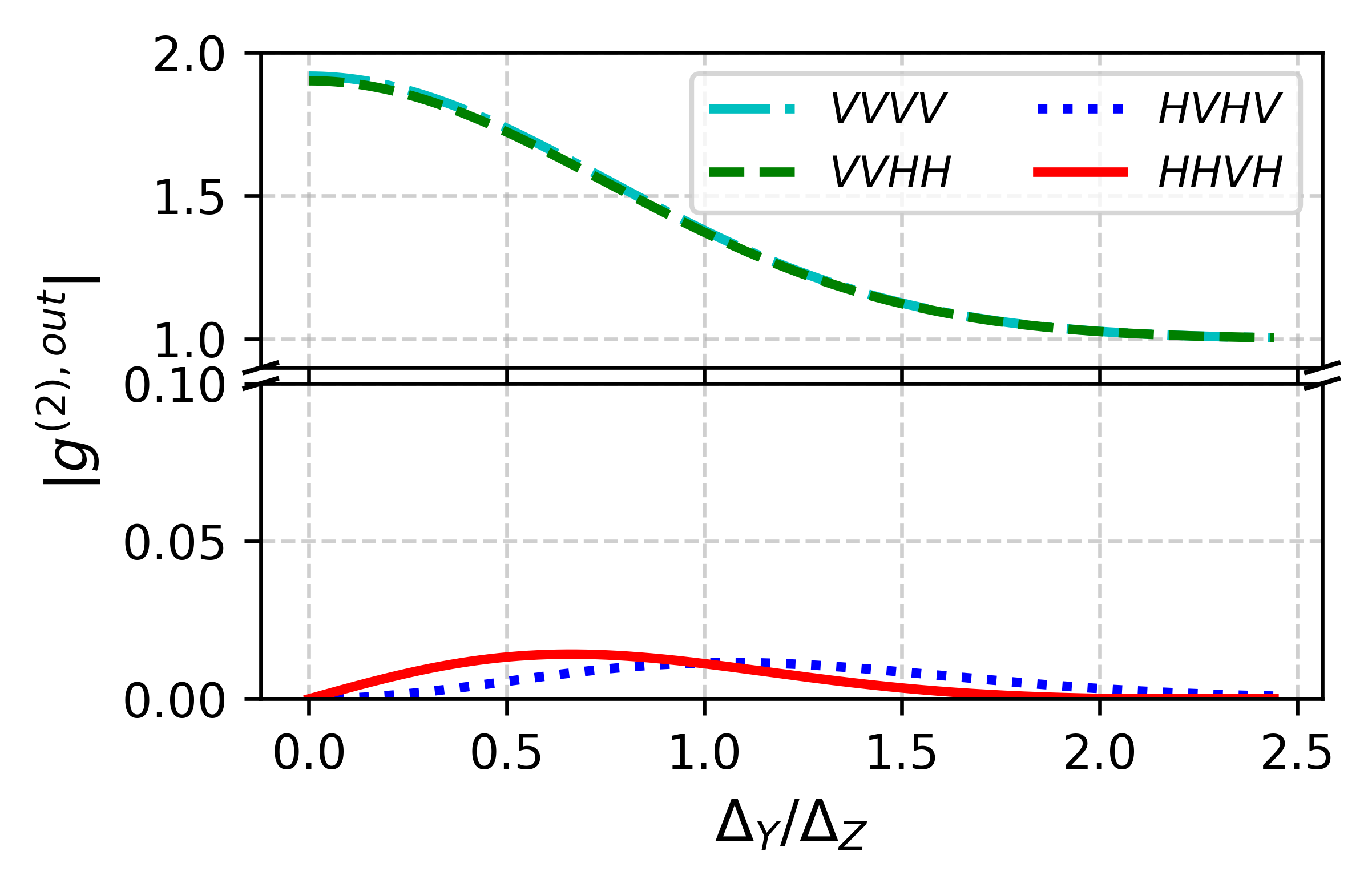}%
	\caption{The second-order coherence as functions of $ \Delta_Y/\Delta_X$ for various post-selected measurements in the far field with $\theta=60^{\circ}$ and $ \Delta_X/\Delta_Z=0.5$. We choose a waist radius of $w_0$=14mm and a wavelength of $\lambda$=8.5mm for the incident Gaussian beam, parameters that are experimentally accessible according to Ref. \cite{iyer2010compact}. Other parameters are the same as those in Fig. 1.}
\end{figure}
\begin{figure*}[!ht]
	\label{Fig.3} \centering \includegraphics[width=1.8\columnwidth]{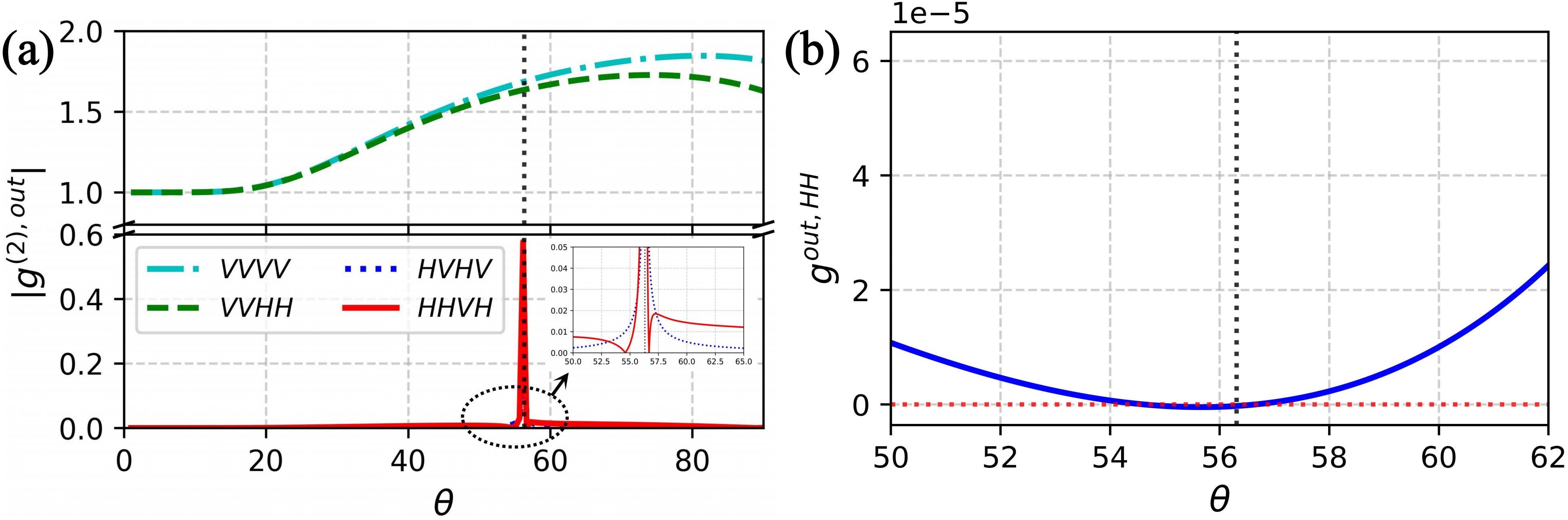}%
	\caption{(a)The second-order coherence $|g^{(2),out}|$ and (b) the  coherence $g^{out,HH}$ as functions of $ \theta$ for various post-selected measurements in the far-field with $ \Delta_X/\Delta_Z=\Delta_Y/\Delta_Z=0.5$. Other parameters are the same as those in Fig. 1.}
\end{figure*}
The two-photon correlations are measured in two configurations: one detector in the vertical (V) mode and the other in the horizontal (H) mode, corresponding to the VVHH component of $|g^{(2),out}|$, and both detectors in the V mode, corresponding to the VVVV component. As shown in Fig. 2, both the VVVV and VVHH components of $|g^{(2),out}|$ decrease with increasing $ \Delta_Y/\Delta_X$, indicating that the coherence between the transmitted and reflected modes become stronger upon propagation. Both components approach 1 for $ \Delta_Y/\Delta_X\approx2$, signifying an uncorrelated state where the two modes become distinguishable. In contrast, as $ \Delta_Y/\Delta_X\to 0$, both components approach a value of 1.9, demonstrating the thermalization of the Gaussian beam in the far-field after its interaction with the dielectric interface, consistent with \cite{you2023multiphoton}.

While point-detector measurements of $|g^{(2),out}_{VVVV}|$ and $|g^{(2),out}_{VVHH}|$ are feasible, a more complete picture is contained in the full field characterization. This richer information, accessible via quantum state tomography \cite{cramer2010efficient,lanyon2017efficient}, reveals more complex effects. The second-order coherences $|g^{(2),out}_{HVHV}|$ and $|g^{(2),out}_{HHVH}|$ exhibit values below one, signifying sub-Poissonian statistics and a photon distribution more confined than that of a coherent state. This suggests potential applicability in sub-shot-noise measurements \cite{agarwal2013quantum}. This sub-Poissonian statistics is achieved and controlled using only post-selection, without the need for complex light-matter interactions \cite{dell2006multiphoton,kondakci2015photonic,folling2005spatial}. The vanishing off-diagonal correlations in the far-field ($|g^{(2),out}_{HVHV}|(0)=|g^{(2),out}_{HHVH}|(0)=0)$, evident from the asymptotic behavior of Eqs. (7) and (10) as $ \Delta_Y/\Delta_Z \to 0$, are consistent with \cite{four-pointcorrelation1,soderholm2001unpolarized}. In contrast, the small but non-zero values of $|g^{(2),out}_{HVHV}|$ and $|g^{(2),out}_{HHVH}|$ elsewhere differ from \cite{you2023multiphoton}, where multi-photon scattering contributes to these correlations. In our case, the vanishing IF shifts for the Gaussian beam suppress this multi-photon contribution, leading to the near-zero values.

The magnitude $|g^{(2),out}_{HHVH}|$, though small at large $ \Delta_Y/\Delta_Z$, does not vanish exactly. Nevertheless, it can be tuned to zero by adjusting the angle of incidence, as illustrated in Fig. 3(a). As shown in Fig. 3(a), $|g^{(2),out}_{HHVH}|$(solid red curve) vanishes on both sides of the Brewster's angle. Recalling Eq. (10), we examine its HH component. Fig. 3(b) shows that this $g^{out,HH}$ component itself disappears at those same incidence angles. 
\begin{figure}[!ht]
	\label{Fig.4} \centering \includegraphics[width=1.02\columnwidth]{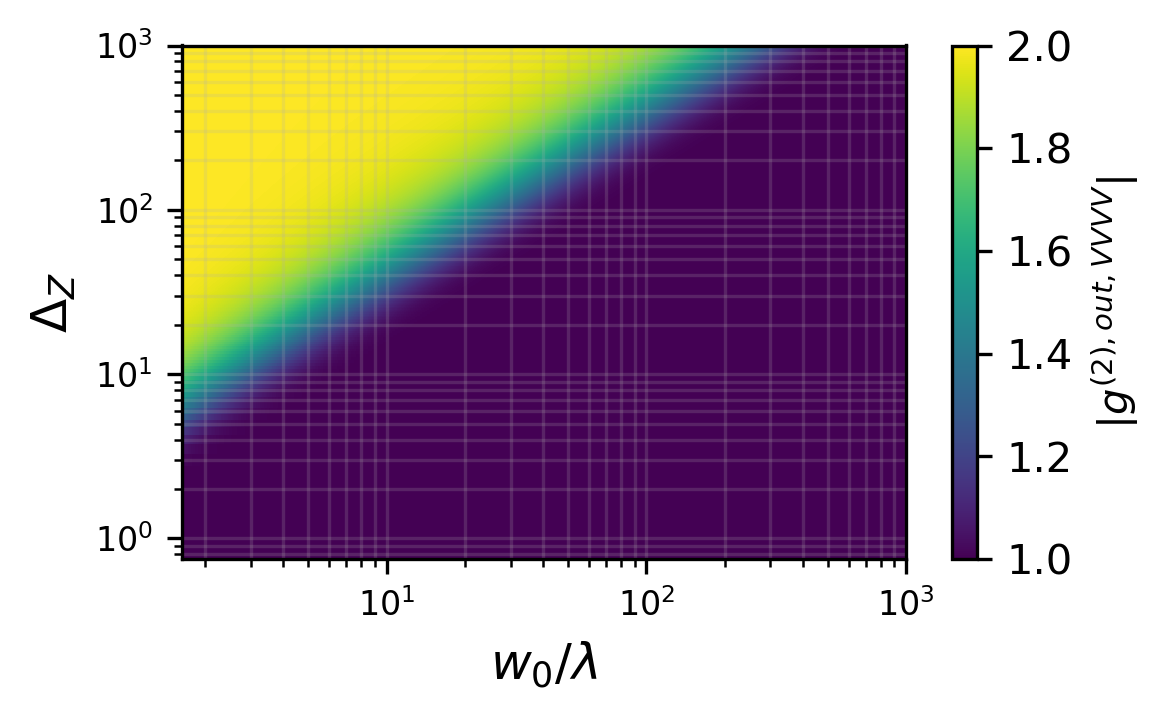}%
	\caption{The VVVV component of the second-order coherence, $|g^{(2),out}_{VVVV}|$, is plotted as a function of $\Delta_Z$ and $ w_0/\lambda$ in the far field, with $\theta=60^{\circ}$ and $\Delta_X/\Delta_Z=\Delta_Y/\Delta_Z=0.5$. A key finding is that $|g^{(2),out}_{VVVV}|$ is independent of the specific value of the waist radius $ w_0$, depending only on the ratio $ w_0/\lambda$. Other parameters are the same as in Fig. 1.}
\end{figure}
This indicates that one H mode in the reflected field makes no contribution to the measurement, leading directly to the vanishing of $|g^{(2),out}_{HHVH}|$. Simultaneously, the HHVH and HVHV components reach their maximum values near the Brewster’s angle. In contrast, the VVVV and VVHH components directly probe the cross-correlation between different polarization modes (reflected and transmitted) at two spatially separated detectors, making it meaningful to introduce an additional degree of freedom for their control. As shown in Fig. 3(b), when the angle of incidence is tuned within $0^{\circ}-20^{\circ}$, $|g^{(2),out}_{VVVV}|$ and $|g^{(2),out}_{VVHH}| $ remain at a value of 1. According to Eqs. (8) and (9), this indicates that these correlations are entirely governed by the indistinguishability term, meaning the two detectors cannot distinguish between the reflected and transmitted spatial modes. As the angle increases beyond this range, $|g^{(2),out}_{VVVV}|$ and $|g^{(2),out}_{VVHH}| $ rise, suggesting that the incidence angle can be used to control the thermalization of the beam in the far field after interacting with the dielectric.

Furthermore, the collimation of the beam is another key factor influencing its thermalization. As shown in Eqs. (7) to (11), the second-order coherence depends only on the ratio of the waist radius $w_{0}$ to the wavelength $\lambda$. Consequently, the specific individual values of $w_{0}$ and $\lambda$ are irrelevant for the subsequent discussion, and only their ratio matters. As shown in Fig. 4, for a fixed $\Delta_Z$, $|g^{(2),out}_{VVVV}|$
decreases from 2 to 1 as the collimation parameter $ w_0/\lambda$ increases. Conversely, for a fixed $ w_0/\lambda$, $|g^{(2),out}_{VVVV}|$ increases from 1 to 2 with propagation distance $\Delta_Z$. This behavior suggests that a poorly collimated Gaussian beam is more susceptible to thermalization upon propagation. This can be understood physically: poor collimation leads to significant beam divergence over distance, causing the reflected and refracted modes to spatially overlap at the detectors, thereby enhancing their mutual correlations and increasing $|g^{(2),out}_{VVVV}|$. In contrast, a well-collimated beam at short distances exhibits $|g^{(2),out}_{VVVV}|\approx1$, signifying an uncorrelated state where the two modes remain distinguishable due to the absence of spatial overlap at the detection plane.
\section{CONCLUSION}
\label{CONCLUSION}
In summary, we have established a framework based on the quantum van Cittert-Zernike theorem to investigate the propagation of quantum coherence in reflected and refracted beams. Our work reveals that the inherent polarization coupling arising from the rotational transformations at a dielectric interface serves as a potent and previously overlooked control knob for manipulating multiphoton correlations and quantum statistics. We demonstrated that the second-order coherence $\left|g^{(2),\mathrm{out}}\right|$ of the outgoing beams is not static but evolves controllably during free-space propagation. Crucially, this control is achieved without conventional light-matter interactions that typically induce decoherence. We identified specific regimes where the measured correlations exhibit sub-Poissonian statistics $\left(\left|g_{HVHV}^{(2),\mathrm{out}}\right|,\left|g_{HHVH}^{(2),\mathrm{out}}\right|<1\right)$, suggesting the potential for realizing fluctuations below the shot-noise level using thermal light. Furthermore, we showed that the incident angle provides a novel degree of freedom to tune these quantum coherence properties, as evidenced by the vanishing of specific correlation components near the Brewster's angle. Another important finding is the role of beam collimation. The quantum coherence depends solely on the ratio $ w_0/\lambda$, implying that poorly collimated beams are more susceptible to thermalization upon propagation due to enhanced spatial overlap of the modes at the detection plane. In contrast, for well-collimated beams, the influence of polarization coupling is negligible, and the reflected and refracted modes remain distinguishable. Our results fundamentally extend the quantum van Cittert-Zernike theorem to scenarios involving beam reflection and refraction, effectively modeling a beam splitter. The ability to manipulate quantum coherence through a naturally accessible parameter like the incident angle, without invoking complex interactions, opens promising avenues for applications in quantum information processing, quantum metrology.
\section{ACKNOWLEDGMENTS}
This work was supported by Natural Science Foundation of Shaanxi Province (Grant No. 2024JC-YBMS-031); Shaanxi Fundamental Science Research Project of Mathematics and Physics (Grant No. 22JSY005); National Natural Science Foundation of China (No. W2411002, 11534008, 91536115);
\appendix

\bibliography{apssamp}
\end{document}